\documentclass[pre,final,twocolumn,showpacs,showkeys]{revtex4}
\usepackage{amsmath}
\usepackage{graphicx}
\usepackage{amsfonts}
\usepackage{amssymb}

\begin{document}

\title{Any-order propagation of the nonlinear Schr{\"o}dinger equation}
\author{Frederick W. Strauch}
\email[Electronic address: ]{frederick.strauch@nist.gov}
\affiliation{National Institute of Standards and Technology, Gaithersburg, Maryland 20899-8423, USA}

\date{\today}

\begin{abstract}
We derive an exact propagation scheme for nonlinear Schr{\"o}dinger equations. This scheme is entirely analogous to the propagation of linear Schr{\"o}dinger equations. We accomplish this by defining a special operator whose algebraic properties ensure the correct propagation. As applications, we provide a simple proof of a recent conjecture regarding higher-order integrators for the Gross-Pitaevskii equation, extend it to multi-component equations, and to a new class of integrators. 
\end{abstract} 

\maketitle

\section{Introduction}

Nonlinear Schr{\"o}dinger equations play a special role in physics. At a fundamental level, it has been suggested that a weak violation of the linearity of quantum mechanics might allow new physics to emerge. Theoretical studies have a long and continuing history \cite{Bialynicki76,Polchinski91,Doebner92}, and have resulted in several precise experimental tests of linearity using neutrons \cite{Shimony79,Shull80,Gaehler81} and ions \cite{Weinberg89,Bollinger89}.  Of course, concrete evidence for such nonlinearity has not been found. At a practical level, nonlinear differential equations have found numerous applications in nonlinear optics, plasma physics, molecular dynamics in biology, and fluid dynamics \cite{Kivshar89}. 

Most spectacularly, the dynamics of the quantum matter wave in dilute Bose-Einstein condensates (BECs) has been described by the time-dependent Gross-Pitaevskii (GP) equation \cite{Dalfovo99}:
\begin{equation}
i \hbar \frac{\partial \Psi({\bf r},t)}{\partial t} = \left(- \frac{\hbar^2 }{2m}\nabla^2 + V_{ext}({\bf r}) + g |\Psi({\bf r},t)|^2 \right) \Psi({\bf r},t).
\label{eq1}
\end{equation}
Among the novel phenomena successfully observed in BECs are four-wave mixing \cite{Deng99}, solitons \cite{Burger99, Denschlag2000, Khaykovich2002}, and vortices \cite{Matthews99, Madison99}.  Each has been accurately modelled by the GP equation (\ref{eq1}), or its multi-component generalizations. 

It is natural to believe that the novel dynamics is a direct result of some intrinsic complexity in the time-evolution. In this paper, we consider the general set of nonlinear Schr{\"o}dinger equations of the form ($\hbar = 1$):
\begin{equation}
i \partial_t \Psi = (H_0 + V_{nl}(\Psi)) \Psi
\label{eq2}
\end{equation}
where the Hamiltonian $H_0$ is a linear, time-independent Hermitian operator, the nonlinear term $V_{nl}(\Psi)$ is a real wave-function dependent (local) potential energy operator, such that the normalization of $\Psi$ is preserved, and we have employed a simplified notation that suppresses the space or spin components of $\Psi$ (see also below).  We find that the exact propagation of (\ref{eq2}) can be accomplished by defining a surprisingly simple nonlinear operator $\hat{V}_{nl}$, and is given by the familiar expression
\begin{equation}
\Psi(t) = \exp\left(-it(H_0 + \hat{V}_{nl})\right) \Psi(0).
\label{eq3}
\end{equation}
Thus, while the resulting evolution of (\ref{eq2}) may be complex, the underlying rules are quite simple.

The motivation for our work is both practical and fundamental. At the practical level, by understanding the underlying structure of nonlinear evolution equations such as (\ref{eq2}) we can design more sophisticated numerical methods. These methods can be much more efficient the more we know about the formal solution. We note two recent examples. 

For real time propagation, it was recently conjectured by Javanainen and Ruostekoski \cite{Java2006} (JR) that all higher-order split-operator algorithms for linear Schr{\"o}dinger equations can be directly applied to the Gross-Pitaevskii (with the same order of accuracy), using only a simple ``most-recent-update'' rule to evaluate the nonlinear potential. We prove this conjecture for all orders, and for multi-component equations in any dimension. 

For imaginary time propagation, standard higher-order split-operator methods can become unstable, due to the presence of negative time-steps.  A new class of fourth-order, positive time-step algorithms has recently been extended to the GP equation in a rotating harmonic trap by Chin and Krotscheck (CK) \cite{Chin2005b}. Using our knowledge of the underlying integration properties, we show how to implement the positive time-step algorithms (in real time) for all trapping potentials.  Physical results using these methods will be presented elsewhere.

At the fundamental level, our work addresses the question: from where does the complexity of nonlinear Schr{\"o}dinger equations arise? Our presentation shows that while the complexity emerges with time, it is {\it not} from the short-time evolution. This is done by bridging the gap between the functional Lie theory of partial differential equations known in the mathematics literature and the operator picture that is more common in the physics literature. This simple picture explicitly demonstrates that there is no extra computational complexity to the propagation of nonlinear equations like the GP equation. 

The structure of this paper is the following.  In Section II we review the propagation law for linear, time-independent Hamiltonians.  In Section III, we present our main result, the propagation law (\ref{eq3}) for nonlinear equations, illustrated by the Gross-Pitaevskii equation.  In Section IV, we prove the propagation law by introducing the Lie operator formalism, and review the Hamiltonian structure of nonlinear Schr{\"o}dinger equations.  In Section V, we study the split-operator methods for real-time propagation, and prove the JR conjecture, and its extension to time-dependent and multi-component equations.  In Section VI, we extend our results to algorithms with positive time steps, and conclude in Section VII.

\section{Linear Propagation}

First, we recall that the linear Schr{\"o}dinger equation ($\hbar = 1$)
\begin{equation}
i \partial_t \Psi = H_0 \Psi 
\label{neweq1}
\end{equation}
with time-independent $H_0$ can be exactly propagated by the scheme
\begin{equation}
\Psi(t) = \exp(-i H_0 t) \Psi(0).\label{neweq2}
\end{equation}
To prove (\ref{neweq2}) we can either take the time derivative to show
that (\ref{neweq1}) is satisfied, or we may use the Taylor series expression
\begin{equation}
\Psi(t) = \sum_{n=0}^{\infty} \frac{t^n}{n!} \left(\partial_t^n\Psi \right)_{t=0}.
\label{neweq3}
\end{equation}
By iterating (\ref{neweq1}) we find that the $n$-th time derivative of $\Psi$ is
\begin{equation}
\partial_t^n \Psi = (-i)^n H_0^n \Psi,
\label{neweq4}
\end{equation}
and thus (\ref{neweq3}) can be evaluated as
\begin{equation}
\Psi(t) = \sum_{n=0}^{\infty} \frac{(-i t)^n}{n!} H_0^n \Psi(0),
\label{neweq5}
\end{equation}
which is precisely what is meant by (\ref{neweq2}).

\section{Nonlinear Propagation}

For the nonlinear Schr{\"o}dinger equation
\begin{equation}
i \partial_t \Psi = (H_0 + V_{nl}(\Psi)) \Psi
\label{neweq6}
\end{equation}
the correct propagation is, at first sight, quite complex.  Recall that in writing (\ref{neweq6}) we use a simplified notation that suppresses not only the spin and spatial coordinates of $\Psi$, but also the fact that the local nonlinear potential $V_{nl}(\Psi) = V_{nl}(\Psi,\Psi^*)$ depends on both $\Psi$ and $\Psi^*$.  In this paper we will consider only real potentials which also satisfy the following reality condition:
\begin{equation}
\frac{\partial V_{nl}}{\partial \Psi} \Psi - \frac{\partial V_{nl}}{\partial \Psi^*} \Psi^* = 0.
\label{neweq6x}
\end{equation}
This is somewhat more general than the restriction that the potential depend only on the local absolute value of $\Psi$, i.e. $V_{nl}(\Psi,\Psi^*) = f(|\Psi|^2)$.  For multi-component equations (to be considered in Section V), where $\Psi$ has components $\phi_n$ and the nonlinear potential has matrix elements $V_{nl,jk}$, this condition generalizes to
\begin{equation}
\sum_{n} \left(\frac{\partial V_{nl,jk}}{\partial \phi_{n}} \phi_n - \frac{\partial V_{nl,jk}}{\partial \phi_{n}^*} \phi_n^*\right) = 0.
\end{equation}

Returning to the propagation of (\ref{neweq6}), we know that a Taylor expansion such as (\ref{neweq3}) still holds:
\begin{equation}
\Psi(t) = \sum_{n=0}^{\infty} \frac{t^n}{n!} \left(\partial_t^n\Psi \right)_{t=0}.
\label{neweq7}
\end{equation}
but a simple form for $\partial_t^n \Psi$ such as (\ref{neweq4}) does not seem readily available. 

If we treat $V(t)=V_{nl}(\Psi(t))$ as a time-dependent potential, we find $\partial_t^n \Psi$ by repeated application of (\ref{neweq6}):
\begin{equation}
\begin{array}{ll}
\partial_t^2 \Psi = & -i (\partial_t V) \Psi - (H_0 + V)^2 \Psi \\
\partial_t^3 \Psi = & -i (\partial_t^2 V) \Psi - 2 (\partial_t V)(H_0 + V) \Psi \\
& - (H_0 + V) (\partial_t V) \Psi + i (H_0 + V)^3 \Psi \\
\partial_t^4 \Psi = & - i (\partial_t^3 V) \Psi - 3 (\partial_t^2 V) (H_0 + V) \Psi - 3 (\partial_t V)^2 \Psi \\
& + 3 i (\partial_t V)(H_0 + V)^2 \Psi - (H_0 + V) (\partial_t^2 V) \Psi \\
& + 2 i (H_0 + V) (\partial_t V) (H_0 + V) \Psi \\
& + i (H_0 + V)^2 (\partial_t V) \Psi + (H_0 + V)^4 \Psi.
\end{array}
\label{neweq7x}
\end{equation}
It appears that the number of terms in $\partial_t^n \Psi$ grows with $n$, by way of the time derivatives of the potential, which must finally be evaluated by differentiating $\Psi$ again. Any hope of achieving an expression as simple as (\ref{neweq4}) appears small.  
 
However, due to the deep Lie structure to the nonlinear Schr{\"o}dinger equation (to be described in the next section), a much simpler formulation can be given. That is, there exists a single {\it nonlinear} operator $\hat{V}_{nl}$ such that
\begin{equation}
\partial_t^n \Psi = (-i)^n (H_0 + \hat{V}_{nl})^n \Psi
\label{neweq11}
\end{equation}
and whose properties we now describe.

By comparing (\ref{neweq6}) to (\ref{neweq11}) with $n=1$, it is clear that
\begin{equation}
\hat{V}_{nl} \Psi = V_{nl}(\Psi) \Psi,
\label{neweq13}
\end{equation}
where we have ``hatted'' the operator on the left to indicate that as an operator it is nonlinear in $\Psi$, while the un-hatted operator on the right can be considered a $\Psi$-dependent linear operator.  
Note also that (\ref{neweq13}) does not uniquely specify $\hat{V}_{nl}$. That is, we must still define the higher powers of $\hat{V}_{nl}$ and its products with $H_0$, each of which occur in the expansion of (\ref{neweq11}). 

For example, a naive application of (\ref{neweq13}) to $H_0 \Psi$ would yield
\begin{equation}
\hat{V}_{nl} H_0 \Psi \overset{?}{=} V_{nl}(H_0 \Psi) H_0 \Psi.
\end{equation}
As indicated by the question mark, this result must be in error, for purely dimensional reasons.  That is, $H_0 \Psi$ has different units than $\Psi$. Thus, any definition should distinguish between $\Psi$ and operators on $\Psi$. Furthermore, our definition should be sufficiently simple in form, yet powerful enough to generate all of the correct terms needed to satisfy (\ref{neweq11}). 

The correct definition for $\hat{V}_{nl}$ is achieved in the following way. Let $K_j$, $j=1, 2, \cdots n$, be an element of some set $\mathcal{K}$ of operators, to be specified. The action of $\hat{V}_{nl}$ is given by the composition rule:
\begin{widetext}
\begin{equation}
\hat{V}_{nl} \left(K_1 \cdots K_n \Psi \right) = i^n \frac{\partial^n}{\partial \lambda_1 \cdots \partial \lambda_n} \left[ V_{nl}(e^{-i \lambda_1 K_1} \cdots e^{-i \lambda_n K_n} \Psi) e^{-i \lambda_1 K_1} \cdots e^{-i \lambda_n K_n} \Psi \right]_{\lambda_1 = \cdots = \lambda_n = 0},
\label{neweq14}
\end{equation}
where we recall that 
\begin{equation}
V_{nl}(e^{-i \lambda_1 K_1} \cdots e^{-i \lambda_n K_n} \Psi) = V_{nl}(e^{-i \lambda_1 K_1} \cdots e^{-i \lambda_n K_n} \Psi, e^{i \lambda_1 K_1} \cdots e^{i \lambda_n K_n} \Psi^*).
\end{equation}
\end{widetext}
Should any $K_j$ itself be nonlinear, its action is defined by using (\ref{neweq14}) recursively. Note that we must also specify the set $\mathcal{K}$. For the nonlinear Schr{\"o}dinger equation, we require that $\mathcal{K} = \{c, H_0, \hat{V}_{nl}\}$ and their linear combinations, where $c$ is any complex scalar. This specification of $\mathcal{K}$ uniquely identifies $H_0^2 = H_0 \times H_0$, which might otherwise be ambiguous.

The definition (\ref{neweq14}) for $\hat{V}_{nl}$ is the main result of this paper. $\hat{V}_{nl}$ is a manifestly nonlinear operator, due to its dependence on $\Psi$ and its algebraic structure.  Nevertheless, it has much in common with linear operators, due to the following remarkable properties. First, the composition rule is dimensionally correct, and commutes with scalar multiplication: 
\begin{equation}
\hat{V}_{nl} c \Psi = c V_{nl}(\Psi) \Psi.
\label{neweq15}
\end{equation}
Note that $c$ here is to be treated as an operator, and thus we use $V_{nl}(e^{-i \lambda c} \Psi) = V_{nl}(e^{-i \lambda c} \Psi, e^{i \lambda c} \Psi^*)$ in (\ref{neweq14}), which by the condition (\ref{neweq6x}) leads to (\ref{neweq15}). Second, it is operator linear:
\begin{equation}
\hat{V}_{nl}(K_1 + K_2)\Psi = \hat{V}_{nl} K_1 \Psi + \hat{V}_{nl} K_2 \Psi.
\label{neweq16}
\end{equation}
Third, $\hat{V}_{nl}$ satisfies the conjugation identity
\begin{widetext}
\begin{equation}
\hat{V}_{nl} e^{-i \lambda_1 K_1} \cdots e^{-i \lambda_n K_n} \Psi = V_{nl}(e^{-i \lambda_1 K_1} \cdots e^{-i \lambda_n K_n} \Psi) e^{-i \lambda_1 K_1} \cdots e^{-i \lambda_n K_n} \Psi.
\label{neweq17}
\end{equation}
\end{widetext}
Conversely, by assuming these properties we could derive the composition rule (\ref{neweq14}).

Finally, note that if the function $V_{nl}$ satisfies $V_{nl} (e^{-i \lambda V_{nl}(\Psi)} \Psi) = V_{nl}(\Psi)$ (which is true for the Gross-Pitaevskii form $V_{nl}(\Psi) = g |\Psi|^2)$, then $\hat{V}_{nl}$ satisfies the power identity
\begin{equation}
\hat{V}_{nl}^n \Psi = \left[ V_{nl}(\Psi) \right]^n \Psi,
\label{neweq18}
\end{equation} 
in which case
\begin{equation}
e^{-i \lambda \hat{V}_{nl}} \Psi = e^{-i \lambda V_{nl}(\Psi)} \Psi.
\label{neweq18a}
\end{equation}

Using the rule (\ref{neweq14}), the exponential propagation law (\ref{eq3}) can be verified by simply taking its time derivative:
\begin{equation}
\begin{array}{ll}
i \partial_t \Psi(t) &= (H_0 + \hat{V}_{nl}) \exp \left(-i t (H_0 + \hat{V}_{nl}) \right) \Psi(0) \\
&= [H_0 + V_{nl}(e^{-i t (H_0 + \hat{V}_{nl})} \Psi(0))] e^{-i t (H_0 + \hat{V}_{nl})} \Psi(0) \\
&= [H_0 + V_{nl}(\Psi(t))] \Psi(t).
\end{array}
\label{neweq18x}
\end{equation}
In deriving (\ref{neweq18x}), we have used the properties of the exponential function, operator linearity (\ref{neweq16}) and the conjugation identity (\ref{neweq17}). 

This derivation of the exponential propagation law, while correct, is only an implicit demonstration that the rule (\ref{neweq14}) generates the higher derivatives $\partial_t^n \Psi$.  An explicit demonstration, for the Gross-Pitaevskii equation, is reproduced in the Appendix, where we show that
\begin{equation}
\Psi(t) = \exp\left(-i t (H_0 + \hat{V}_{nl})\right) \Psi(0) + O(t^5).
\label{neweq30}
\end{equation}
In the next section we will provide an independent argument to show that this is correct to any order in $t$ and for any nonlinear Schr{\"o}dinger equation.

\section{Lie Operator Proof}

To explicitly prove (\ref{neweq11}) for general $n$, and therefore the exact exponential evolution (\ref{eq3}), we use the Lie operator formalism, which easily handles nonlinear evolution equations, and has found great use in the mathematical study of differential equations \cite{OlverBook,SanzSerna97}. In fact, all linear and many nonlinear Schr{\"o}dinger equations can be written as the Hamiltonian evolution of a classical field \cite{Strocchi66,Heslot85,Bialynicki76,Weinberg89b,Jones92}, and this evolution (in an infinite-dimensional phase space) can in turn be described using Lie operators \cite{Dragt76,Dragt83}. These formal methods will justify the presentation of the previous section.  While there have been other general studies of nonlinear wave equations using Lie operators \cite{McLachlan94, Herbst94, Rouhi95, Lanser99}, the new approach developed here directly shows how to convert the Lie {\it differential} structure into a novel {\it algebraic} structure, a nontrivial conceptual shift that can simplify many calculations.

The pair of differential equations
\begin{equation}
\begin{array}{ll}
\partial_t \Psi & = -i(H_0 +  V_{nl}(\Psi)) \Psi \\
\partial_t \Psi^* & = +i(H_0 +  V_{nl}(\Psi)) \Psi^*		
\end{array}
\label{neweq31}
\end{equation}
(note that we consider $V_{nl}(\Psi^*) = V_{nl}^*(\Psi) = V_{nl}(\Psi)$) can be succinctly written as
\begin{equation}
\begin{array}{ll}
\partial_t \Psi &= \mathcal{L} \Psi \\
\partial_t \Psi^* &= \mathcal{L} \Psi^*
\end{array}
\label{neweq32}
\end{equation}
where the Lie operators $\mathcal{L} = \mathcal{L}_H + \mathcal{L}_V$ are functional derivative operators, defined by 
\begin{equation}
\mathcal{L}_H = -i\left( H_0 \Psi \frac{\partial}{\partial \Psi} - H_0 \Psi^* \frac{\partial}{\partial \Psi^*} \right),
\label{neweq33}
\end{equation}
\begin{equation}
\mathcal{L}_V = -i\left( V_{nl}(\Psi) \Psi \frac{\partial}{\partial \Psi} - V_{nl}(\Psi) \Psi^* \frac{\partial}{\partial \Psi^*} \right).
\label{neweq34}
\end{equation}

The Lie operators (\ref{neweq33}) and (\ref{neweq34}) arise naturally by treating $\Psi$ as a classical field. Note that, in coordinates, (\ref{neweq33}) represents
\begin{equation}
\mathcal{L}_H = -i\int d^3 {\bf r} \left( (H_0\Psi)({\bf r}) \frac{\delta}{\delta \Psi({\bf r})} -  (H_0 \Psi^*)({\bf r}) \frac{\delta}{\delta \Psi^*({\bf r})} \right),
\label{neweq34x}
\end{equation}
with a similar expression for (\ref{neweq34}). The form of this expression can be understood by introducing the Poisson bracket \cite{Strocchi66}
\begin{equation}
\{\mathcal{A},\mathcal{B}\} = -i \int d^3 {\bf r} \left( \frac{\delta \mathcal{A}}{\delta \Psi({\bf r})} \frac{\delta \mathcal{B}}{\delta \Psi^*({\bf r})} - \frac{\delta \mathcal{B}}{\delta \Psi({\bf r})} \frac{\delta \mathcal{A}}{\delta \Psi^*({\bf r})} \right),
\label{neweq34y}
\end{equation}
where $\mathcal{A}$ and $\mathcal{B}$ are functionals of $\Psi({\bf r})$. This bracket introduces a symplectic (i.e. canonical) structure on a phase space where $\Psi({\bf r})$ and $\Psi^*({\bf r})$ play the role of coordinates and momenta. The dynamical evolution of any phase-space function(al) is generated by a Hamiltonian functional, which for the Gross-Pitaevskii equation is
\begin{equation}
\mathcal{H} = \int d^3 {\bf r} \left(\Psi^*({\bf r}) H_0 \Psi({\bf r}) + \frac{1}{2} g |\Psi({\bf r})|^4 \right).
\label{neweq34z}
\end{equation}
For example, the time evolution of $\Psi({\bf r})$ is given by
\begin{equation}
\partial_t \Psi({\bf r}) = \{\Psi({\bf r}),\mathcal{H}\} = \mathcal{L} \Psi({\bf r}),
\end{equation}
where we implicitly defined the Lie operator, i.e. $\mathcal{L} = \{\cdot,\mathcal{H}\} = -\{\mathcal{H}, \cdot\}$. Substituting $\mathcal{H}$ of (\ref{neweq34z}) in the bracket (\ref{neweq34y}) will produce the Lie operators $\mathcal{L}_H$ (\ref{neweq33}) and $\mathcal{L}_V$ (\ref{neweq34}) with $V_{nl}(\Psi) = g |\Psi|^2$.  

This quick review shows how nonlinear Schr{\"o}dinger equations can be understood as the Hamiltonian dynamics of a classical field. Furthermore, linear Schr{\"o}dinger equations correspond to Hamiltonians which are quadratic (or bilinear) in the field variables.  We now show that this Hamiltonian dynamics can be represented by an exponential propagation law.

First, we find that $\partial_t^n \Psi$ can be elegantly written in terms of $\mathcal{L}=\mathcal{L}_H + \mathcal{L}_V$:
\begin{equation}
\partial_t^n \Psi = (\mathcal{L}_H + \mathcal{L}_V)^n \Psi.
\label{neweq35}
\end{equation}
This is proven by induction.  That is, we write the right-hand-side of (\ref{neweq35}) as some function $F_n(\Psi,\Psi^*)$, which we assume is $\mathcal{L}^n \Psi$.  We then compute
\begin{widetext}
\begin{equation}
\begin{array}{ll}
F_{n+1}(\Psi,\Psi^*) & = \partial_t F_n (\Psi,\Psi^*) \\
& = \partial_{\Psi} F_n(\Psi,\Psi^*) \partial_t \Psi + \partial_{\Psi^*} F_n(\Psi,\Psi^*)\partial_t \Psi^* \\
& = -i (H_0 \Psi + V_{nl}(\Psi)\Psi) \partial_{\Psi} F_n(\Psi,\Psi^*) + i (H_0 \Psi^* + V_{nl}(\Psi) \Psi^*) \partial_{\Psi^*} F_n(\Psi,\Psi^*) \\
& = \mathcal{L} F_n(\Psi,\Psi^*) \\
& = \mathcal{L}^{n+1} \Psi.
\end{array}
\label{neweq35x}
\end{equation}
\end{widetext}
Then, using (\ref{neweq35}) in the Taylor expansion (\ref{neweq7}), we find 
\begin{equation}
\Psi(t) = \left( \exp( t \mathcal{L}_H + t \mathcal{L}_V ) \Psi \right)_{\Psi = \Psi(0)}.
\label{neweq36}
\end{equation}

This demonstration in fact directly implies our earlier claim (\ref{eq3}), but to show this we must convert the Lie operators $\mathcal{L}_H$ and $\mathcal{L}_V$ into the original operators $H_0$ and $\hat{V}_{nl}$.  We recall a key property of Lie operators: for any function $f(\Psi)$, we have \cite{Dragt76, Dragt83}
\begin{equation}
e^{\lambda \mathcal{L}} f(\Psi) = f\left(e^{\lambda \mathcal{L}} \Psi\right).
\label{neweq37}
\end{equation}
We also define a set of Lie operators corresponding to our original set $\mathcal{K}$ of Hermitian operators, i.e.
\begin{equation}
\mathcal{L}_j = -i \left( K_j \Psi \frac{\partial}{\partial \Psi} - K_j \Psi^* \frac{\partial}{\partial \Psi^*} \right).
\label{neweq38}
\end{equation}
Then, since \textit{all} Lie operators act linearly, we have the identity
\begin{equation}
\mathcal{L}_1 \cdots \mathcal{L}_n \mathcal{L}_V \Psi = \frac{\partial^n}{\partial \lambda_1 \cdots \lambda_n} \left( e^{\lambda_1 \mathcal{L}_1} \cdots e^{\lambda_n \mathcal{L}_n} \mathcal{L}_V \Psi \right)_{\lambda_1 = \cdots = \lambda_n = 0}
\label{neweq39}
\end{equation}

To proceed, consider the case $n=2$ with each $K_j$ a linear operator. In this case we have $\mathcal{L}_j^n \Psi = (-i)^n K_j^n \Psi$.  In addition, $\mathcal{L}_V \Psi = V_{nl}(\Psi) \Psi$, and using (\ref{neweq37}) we compute
\begin{widetext}
\begin{equation}
\begin{array}{ll}
e^{\lambda_1 \mathcal{L}_1} e^{\lambda_2 \mathcal{L}_2} \mathcal{L}_V \Psi & = e^{\lambda_1 \mathcal{L}_1} e^{\lambda_2 \mathcal{L}_2} V_{nl}(\Psi) \Psi \\
& = e^{\lambda_1 \mathcal{L}_1} V_{nl}(e^{\lambda_2 \mathcal{L}_2} \Psi) e^{\lambda_2 \mathcal{L}_2} \Psi \\
& = e^{\lambda_1 \mathcal{L}_1} V_{nl}(e^{-i \lambda_2 K_2} \Psi) e^{-i \lambda_2 K_2} \Psi \\
& = V_{nl}(e^{-i \lambda_2 K_2} e^{\lambda_1 \mathcal{L}_1} \Psi) e^{-i \lambda_2 K_2} e^{\lambda_1 \mathcal{L}_1} \Psi \\
& = V_{nl}(e^{-i \lambda_2 K_2} e^{-i \lambda_1 K_1} \Psi) e^{-i \lambda_2 K_2} e^{-i \lambda_1 K_1} \Psi.
\end{array}
\label{neweq40}
\end{equation}
\end{widetext}
Thus, the Lie operators $\mathcal{L}_j$ act from left-to-right, while the linear operators $K_j$ act from right-to-left. Pursuing the calculation to general $n$ we find
\begin{widetext}
\begin{equation}
e^{\lambda_1 \mathcal{L}_1} \cdots e^{\lambda_n \mathcal{L}_n} \mathcal{L}_V \Psi = V_{nl}(e^{-i \lambda_n K_n} \cdots e^{-i \lambda_1 K_1} \Psi) e^{-i \lambda_n K_n} \cdots e^{-i \lambda_1 K_1} \Psi.
\label{neweq40x}
\end{equation}
\end{widetext}

Substituting (\ref{neweq40x}) in (\ref{neweq39}) and comparing with (\ref{neweq14}), we see that we should make the following identification:
\begin{equation}
\mathcal{L}_1 \cdots \mathcal{L}_n \mathcal{L}_V \Psi = (-i)^n \hat{V}_{nl} K_n \cdots K_1 \Psi.
\label{neweq41}
\end{equation}
Finally, we observe that our assumption of linear $K_j$ was not essential and can be dropped, so long as we recursively use (\ref{neweq14}).  Using this correspondence we can now conclude that (\ref{neweq11}) and (\ref{neweq35})are equivalent.

To summarize, we have shown how the Lie formalism directly yields a simple expression for the $\partial_t^n \Psi$ (\ref{neweq35}) and exponential propagation (\ref{neweq36}). We have also converted the formal Lie products into a simple operator expression by (\ref{neweq41}). Thus, we can associate every term in the Lie operator expansion of (\ref{neweq36}) with every term in the expansion of (\ref{eq3}). Each term is found to be identical using the rule given by (\ref{neweq14}). Altogether, we have shown how the nonlinear Schr{\"o}dinger equation can be exactly propagated, to all orders in $t$, by a simple exponential.

\section{Split-Operator Algorithms}

Recently, Javanainen and Ruostekoski (JR) have studied the properties of higher-order split-operator approaches to the Gross-Pitaevskii equation \cite{Java2006}.  A split-operator scheme approximates an exponential operator $e^{\lambda (A+B)}$ by a product of the (hopefully) simpler individual exponentials, $e^{\lambda A}$ and $e^{\lambda B}$, in the following way:
\begin{equation}
e^{\lambda (A+B)} = \prod_{k=1}^{N_n} e^{\lambda a_k A} e^{\lambda b_k B} + O(\lambda^{n+1}),
\label{neweq42}
\end{equation}
where the coefficients $a_k$ and $b_k$ and the number $N_n$ have been chosen to achieve the specified order of accuracy. Such an approximation is known as an $n$-th order splitting scheme. These schemes have been extensively studied in the case that $A$ and $B$ are linear operators, and many properties of this expansion have been established. A review of splitting methods can be found in \cite{McLachlan2002}.

Much less familiar is the case when either $A$ or $B$ is a nonlinear operator. It was observed in \cite{Java2006} that the standard split-operator approach works for the one-dimensional Gross-Pitaevskii equation with $A = -i H_0$ and $B = - i V_{nl}(\Psi) = -i g|\Psi|^2$, if the most recent update of the wavefunciton is used to evaluate $B$.  For example, the second-order splitting (due to Strang \cite{Strang68}) is
\begin{equation}
e^{\lambda (A+B)} = e^{\lambda A/2} e^{\lambda B} e^{\lambda A/2} + O(\lambda^3).
\label{neweq43}
\end{equation}
To use this to propagate the Gross-Pitaevskii equation, the correct rule is
\begin{equation}
\Psi(t+\tau) = e^{-i \tau H_0/2} e^{-i \tau V_{nl}(\Phi_1)} e^{-i \tau H_0/2} \Psi(t) + O(\tau^3),
\label{neweq44}
\end{equation}
where $\Phi_1 = \exp(-i \tau H_0 /2) \Psi(t)$ is the most recent wavefunction.  JR found that this prescription will also work for the fourth-order splitting (due to Forest and Ruth \cite{Forest90} and others \cite{Creutz89,Candy91,Yoshida90,Suzuki90}):  
\begin{widetext}
\begin{equation}
e^{\lambda (A+B)} = e^{\lambda s A/2} e^{\lambda s B} e^{\lambda (1-s) A/2} e^{\lambda (1-2s) B} e^{\lambda (1-s) A/2} e^{\lambda s B} e^{s \lambda A/2} + O(\lambda^5)
\label{neweq45}
\end{equation}
\end{widetext}
with $s = 1/(2 - 2^{1/3})$, and conjectured that this was true for all higher-order algorithms (note that fourth-order schemes for nonlinear equations were previously studied in \cite{Bandrauk94}).

The results of the previous section show that this conjecture is indeed true for all split-operator schemes, for (nearly) all nonlinear Schr{\"o}dinger equations, and for multi-component wavefunctions of any dimension. That is, we have shown that the exact time-propagation of the GP equation is an exact exponential. While $\hat{V}_{nl}$ is a nonlinear operator, it satisfies the important property of operator linearity. Thus, any split-operator scheme for linear operators can be directly applied to the nonlinear Schr{\"o}dinger equation.  These propagation schemes are particularly simple should the nonlinear operator satisfy the power identity (\ref{neweq18}).

The explicit argument is the following. We have proven the exponential propagation law
\begin{equation}
\Psi(t+\tau) = e^{-i \tau (H_0 + \hat{V}_{nl})} \Psi(t).
\label{neweq45x}
\end{equation}
Then, using (\ref{neweq42}) with $\lambda = -i \tau$, $A = H_0$ and $B = \hat{V}_{nl}$ we find
\begin{equation}
\Psi(t+\tau) = \prod_{k=1}^{N_n} e^{-i \tau a_k H_0} e^{-i \tau b_k \hat{V}_{nl}} \Psi(t) + O(\tau^{n+1}).
\label{neweq46a}
\end{equation}
To evaluate the exponentials, we use the conjugation (\ref{neweq17}) and power (\ref{neweq18}) identities, such
that 
\begin{equation}
e^{-i \tau b_k \hat{V}_{nl}} U_k \Psi(t) = e^{-i \tau b_k V(U_k \Psi(t))} U_k \Psi(t)
\label{neweq46x}
\end{equation}
where
\begin{equation}
U_k = \prod_{j <k} e^{-i \tau a_k H_0} e^{-i \tau b_k \hat{V}_{nl}}.
\end{equation}
By iterating (\ref{neweq46x}) down to $k=1$, we find that the split-operator scheme (\ref{neweq46a}) becomes
\begin{equation}
\Psi(t+\tau) = \prod_{k=1}^{N_n} e^{-i \tau a_k H_0} e^{-i \tau b_k V_{nl}(\Phi_k)} \Psi(t) + O(\tau^{n+1}),
\label{neweq46}
\end{equation}
where $\Phi_k$ is an auxiliary wavefunction given by the ``most-recent-update'' rule:
\begin{equation}
\Phi_k = \prod_{j < k} e^{-i \tau a_j H_0} e^{-i \tau b_j V_{nl}(\Phi_j)} \Psi(t).
\label{neweq47}
\end{equation}

Note that no assumptions regarding the dimensionalities of $H_0$, $V_{nl}$, and $\Psi$ have been introduced. The only property that is somewhat special to the GP equation is the power identity (\ref{neweq18}). Any nonlinear Schr{\"o}dinger equation satisfying this identity can be easily integrated using the split-operator scheme (\ref{neweq46})-(\ref{neweq47}).  Previous split-operator approaches to the nonlinear Schr{\"o}dinger equation \cite{Taha84b,Weideman86,Pathria90}, have typically used only low order splittings such as (\ref{neweq43}). The first important studies of higher-order methods include Bandrauk and Shen \cite{Bandrauk94}, who treat the nonlinearity as a time-dependent potential, and McLachlan \cite{McLachlan94} who applied Lie operator methods more generally.

To highlight the generality of this approach, we apply it to two multi-component examples.  First, consider the coupled nonlinear Schr{\"o}dinger equations studied in \cite{Bandrauk94}:
\begin{equation}
\begin{array}{ll}
i \partial_t \phi_1 &= \left(H_1 + g_{11} |\phi_1|^2 + g_{12} |\phi_2|^2\right) \phi_1 \\
i \partial_t \phi_2 &= \left(H_2 + g_{21} |\phi_1|^2 + g_{22} |\phi_2|^2\right) \phi_2.
\end{array} 
\end{equation}
By treating $\phi_1$ and $\phi_2$ as elements of a two-component wavefunction $\Psi$, this equation takes the same form as (\ref{eq2}) but where $H_0$ and $V_{nl}$ are now matrices:
\begin{equation}
H_0 = \left(\begin{array}{cc} H_1 & 0 \\ 0 & H_2 \end{array} \right),
\label{nlse2a}
\end{equation}
and
\begin{equation}
V_{nl}(\Psi) = \left(\begin{array}{cc} g_{11} |\phi_1|^2 + g_{12} |\phi_2|^2 & 0 \\
0 & g_{21} |\phi_1|^2 + g_{22} |\phi_2|^2 \end{array} \right).
\label{nlse2b}
\end{equation}
One can easily show that the corresponding nonlinear operator $\hat{V}_{nl}$ satisfies the power identity (\ref{neweq18}), and therefore the integration scheme presented above can be implemented directly, since each matrix is diagonal.

A nontrivial example is the following four-component equation:
\begin{equation}
\begin{array}{ll}
i \partial_t \phi_1 &= (H_1 + U_1) \phi_1 + 2 g \phi_2^* \phi_3 \phi_4 \\
i \partial_t \phi_2 &= (H_2 + U_2) \phi_2 + 2 g \phi_1^* \phi_3 \phi_4 \\
i \partial_t \phi_3 &= (H_3 + U_3) \phi_3 + 2 g \phi_4^* \phi_1 \phi_2 \\
i \partial_t \phi_4 &= (H_4 + U_4) \phi_4 + 2 g \phi_3^* \phi_1 \phi_2.
\end{array}
\end{equation}
Equations of this form arise in four-wave mixing in BECs \cite{Deng99,Trippenbach2000}.  Note that the potentials $U_j$ can depend on $\phi$, but take forms similar to previous example; they are known as the phase-modulation terms, while the mixing terms have been written out explicitly.  A direction matrix formulation of this equation does not obviously lead to an operator $\hat{V}_{nl}$ that satisfies the power identity (\ref{neweq18}).  However, one can split the nonlinear interactions into separate non-commuting terms which individually satisfy the identity.  That is, an equivalent evolution equation reads
\begin{equation}
i \partial_t \Psi = \left(H_0 + \hat{V}_{0} + \hat{V}_{1} + \hat{V}_{2} + \hat{V}_{3} + \hat{V}_{4} \right) \Psi
\end{equation}
where $H_0$ and $V_0(\Psi)$ are matrices similar to (\ref{nlse2a}) and (\ref{nlse2b}), and 
\begin{equation}
V_{1}(\Psi) = g \left(\begin{array}{cccc} 0 & 0 & \phi_2^* \phi_4 & 0 \\
0 & 0 & 0 & 0 \\
\phi_4^* \phi_2 & 0 & 0 & 0 \\
0 & 0 & 0 & 0 \end{array}\right),
\end{equation}
\begin{equation}
V_{2}(\Psi) = g \left(\begin{array}{cccc} 0 & 0 & 0 & \phi_2^* \phi_3 \\
0 & 0 & 0 & 0 \\
0 & 0 & 0 & 0 \\
\phi_3^* \phi_2 & 0 & 0 & 0 \end{array}\right),
\end{equation}
\begin{equation}
V_{3}(\Psi) = g \left(\begin{array}{cccc} 0 & 0 & 0 & 0 \\
0 & 0 & \phi_1^* \phi_4 & 0 \\
0 & \phi_4^* \phi_1 & 0 & 0 \\
0 & 0 & 0 & 0 \end{array}\right),
\end{equation}
\begin{equation}
V_{4}(\Psi) = g \left(\begin{array}{cccc} 0 & 0 & 0 & 0 \\
0 & 0 & 0 & \phi_1^* \phi_3 \\
0 & 0 & 0 & 0 \\
0 & \phi_3^* \phi_1 & 0 & 0 \end{array}\right).
\end{equation}
Each of these matrices represents a rotation between two components of $\Psi$, with a rotation angle determined by the other two components (which remain unchanged).  As nonlinear operators, each satisfies the power identity (\ref{neweq18}) and thus their exponential has a simple implementation.  Therefore, an integration routine for this set of equations can be immediately implemented, using simple generalizations of splitting formulae such as (\ref{neweq42}), e.g.
\begin{equation}
\begin{array}{ll}
\Psi(t+\tau) = & e^{-i \tau H_0/2} e^{-i \tau \hat{V}_0/2} e^{-i \tau \hat{V}_1/2} e^{-i \tau \hat{V}_2/2} \\ 
& \times e^{-i \tau \hat{V}_3/2} e^{-i \tau \hat{V}_4} e^{-i \tau \hat{V}_3/2} e^{-i \tau \hat{V}_2/2} \\
& \times e^{-i \tau \hat{V}_1/2} e^{-i \tau \hat{V}_0/2} e^{-i \tau H_0/2} \Psi(t) + O(\tau^3).
\end{array}
\end{equation}
 
Finally we observe that this approach treats the nonlinear operator $\hat{V}_{nl}$ as a time-{\textit{independent}} operator. For problems which have an {\it explicit} time-dependence (but not through $\Psi$), one can use standard sequencing techniques (as indicated in \cite{Forest90}, see also \cite{Chin2002}) to evaluate the operators. However, as observed by JR, if $V_{nl}$ is treated as a time-dependent operator, such sequencing does not produce expressions as simple as (\ref{neweq46})-(\ref{neweq47}).  

\section{Positive Time-Step Algorithms}

A peculiarity of the general splitting (\ref{neweq42}) is that, for methods of third-order or higher, negative time-steps must appear, i.e. one of each of the coefficients $\{a_j\}$ and $\{b_j\}$ must be negative \cite{Sheng89,Suzuki91,Goldman96}. These negative time-steps have two other negative effects: they introduce potential instability in imaginary time evolution, and generate large error terms. By introducing the new operator $[B,[B,A]]$ into the product can one find positive time-step algorithms of third and fourth order \cite{Suzuki95,Chin97}, which typically have nicer errors than schemes like Forest-Ruth \cite{Chin2002}.  One especially simple product is 
\begin{equation}
e^{\lambda (A+B)} = e^{\lambda B/6} e^{\lambda A/2} e^{\lambda 2 \tilde{B} /3} e^{\lambda A/2} e^{\lambda B/6} + O(\lambda^5)
\label{chinprod1}
\end{equation}
with 
\begin{equation}
\tilde{B} = B + \frac{1}{48} \lambda^2 [B,[B,A]].
\label{chinprod2}
\end{equation}
Recently Chin and Krotscheck (CK) \cite{Chin2005b} have shown how to use this factorization for imaginary-time propagation of Bose-Einstein condensates in a rotating harmonic trap.  In their approach, $V_{nl}$ is treated as a time-dependent potential whose action is evaluated by a self-consistent iterative procedure.

Letting $A = - \nabla^2$, i.e. the kinetic energy operator (with $\hbar = 2m = 1$), and $B = (V_{ext} + \hat{V}_{nl})$, and using the composition rule (\ref{neweq14}), we find that the relevant commutator for the Gross-Pitaevskii equation, when acting on $\Psi$, is 
\begin{equation}
\begin{array}{ll}
[B, [B, A]] \to& g^2 (\nabla |\Psi|^2) \cdot (\nabla |\Psi|^2) + 2 g^2 |\Psi|^2 \nabla^2 |\Psi|^2 \\ 
& + \ 2 g |\Psi|^2 \nabla^2 V_{ext} - (\nabla V_{ext}) \cdot (\nabla V_{ext}).
\end{array}
\end{equation}
Using this expression in (\ref{chinprod1}) and (\ref{chinprod2}), we can again construct an integrator using the ``most-recent-update'' rule just as in the previous section.  Furthermore, this approach is somewhat simpler than the self-consistent iteration method used by CK and can be applied to any potential, though at the cost of computing additional derivatives.  

There is an important distinction, however, between this approach and the methods developed by CK.  The approach presented here is not directly applicable to imaginary-time propagation.  That is, simply considering the transformation $\tau = i t$ in (\ref{eq2}) leads to an equation that does not preserve the norm of the wavefunction.  While an exact exponential propagation law such as (\ref{eq3}) for such an equation does exist, it requires a composition rule slightly different from (\ref{neweq14}).  In addition, what is really implemented by CK's method is an integrator for the following nonlinear and {\it{nonlocal}} equation:
\begin{equation}
\frac{\partial \Psi({\bf r},\tau)}{\partial \tau} = - \left(- \frac{\hbar^2 \nabla^2}{2m} + V_{ext}({\bf r}) + \mathcal{N}(\Psi,{\bf r})\right) \Psi({\bf r},\tau)
\end{equation}
with
\begin{equation}
\mathcal{N}(\Psi,{\bf r}) = g \frac{|\Psi({\bf r},\tau)|^2}{\int_{-\infty}^{\infty} d^3{\bf r'} |\Psi({\bf r'},\tau)|^2}.
\end{equation}
Development of explicit integration schemes for this operator is left for the future.  

\section{Conclusion}

In this paper we have presented a new approach to the propagation of nonlinear Schr{\"o}dinger equations.  By translating the functional Lie theory of differential equations to a nonlinear operator, we have shown how to encode the propagation law as an exact exponential, just like linear Schr{\"o}dinger equations.  This formalism was then applied to the construction of higher-order split-operator methods.  In particular, the ``most-recent-update'' conjecture was proven to any order, and new methods were proposed for multi-component equations arising in nonlinear pulse propagation and four-wave mixing of BECs.  Finally, it was shown how these methods could simplify the implementation of a more efficient fourth-order algorithm for real-time propagation in arbitrary potentials (albeit not for imaginary-time).  As the propagation of nonlinear Schr{\"o}dinger equations remains an important topic for many physical applications, we believe this approach can be a productive tool not only for numerical methods but also for analytical approximations; applications to wave-mixing in BECs will be presented elsewhere.

\begin{acknowledgments}
We thank P. R. Johnson, E. Tiesinga, C. Clark, and B. Schneider for stimulating discussions and helpful advice, and acknowledge support from the NRC.

\end{acknowledgments}

\appendix*
\section{}

In this Appendix we show how the rule (\ref{neweq14}) correctly generates the higher derivatives $\partial_t^n \Psi$ via
\begin{equation}
\partial_t^n \Psi = (-i)^n (H_0 + \hat{V}_{nl})^n \Psi,
\label{appeq1}
\end{equation}
for $n=1$ to $n=4$.  We consider the important example of the Gross-Pitaevskii equation, which has $V_{nl}(\Psi) = g |\Psi|^2$ (recall that we have suppressed the position dependence). The $\partial_t^n \Psi$ of (\ref{neweq7x}) take the complex form:
\begin{widetext}
\begin{equation}
\partial_t^2 \Psi = 
- H_0^2 \Psi - g H_0(|\Psi|^2 \Psi) - 2 g |\Psi|^2 H_0 \Psi + g \Psi^2 H_0 \Psi^* - g^2 |\Psi|^4 \Psi,
\label{appeq2}
\end{equation}
\begin{equation}
\begin{array}{ll}
\partial_t^3 \Psi  = & i H_0^3 \Psi + i g H_0^2(|\Psi|^2 \Psi) + 2 i g H_0 (|\Psi|^2 H_0 \Psi) -i g H_0 (\Psi^2 H_0 \Psi^*) \\ 
& + 2 i g |\Psi|^2 H_0^2 \Psi + 2 i g \Psi^*(H_0 \Psi)^2 - 4 i g \Psi (H_0 \Psi) (H_0 \Psi^*) + i g \Psi^2 H_0^2 \Psi^* \\
& + i g^2 H_0(|\Psi|^4 \Psi) + 2 i g^2 |\Psi|^2 H_0(|\Psi|^2 \Psi) + 3 i g^2 |\Psi|^4 (H_0 \Psi)\\
& - 4 i g^2 |\Psi|^2 \Psi^2 (H_0 \Psi^*) + i g^2 \Psi^2 H_0 (|\Psi|^2 \Psi^*) + i g^3 |\Psi|^6 \Psi,
\end{array}
\label{appeq3}
\end{equation}
\begin{equation}
\begin{array}{ll}
\partial_t^4 \Psi = & H_0^4 \Psi + g H_0^3(|\Psi|^2 \Psi) + 2 g H_0^2(|\Psi|^2 H_0 \Psi) - g H_0^2(\Psi^2 H_0 \Psi^*) \\
& + 2 g H_0 (|\Psi|^2 H_0 \Psi) + 2 g H_0(\Psi^*(H_0 \Psi)^2) - 4 g H_0 (\Psi (H_0 \Psi)(H_0 \Psi^*)) \\
& + g H_0(\Psi^2 H_0^2 \Psi^*) + 2 g |\Psi|^2 H_0^3 \Psi + 6 g \Psi^*(H_0 \Psi) (H_0^2 \Psi) \\
& - 6 g (H_0 \Psi)^2 (H_0 \Psi^*) - 6 g \Psi (H_0 \Psi^*) H_0^2 \Psi + 6 g \Psi (H_0 \Psi) H_0^2 \Psi^* \\
& - g \Psi^2 H_0^3 \Psi^* + g^2 H_0^2(|\Psi|^4 \Psi) + 2 g^2 H_0(|\Psi|^2 H_0(|\Psi|^2 \Psi)) \\
& + g^2 H_0(\Psi^2 H_0(|\Psi|^2 \Psi^*)) + 6 g^2 \Psi^*(H_0 \Psi)H_0 (|\Psi|^2 \Psi) - 6 g^2 \Psi (H_0 \Psi^*) H_0 (|\Psi|^2 \Psi) \\
& + 6 g^2 \Psi (H_0 \Psi) H_0 (|\Psi|^2 \Psi^*) + 6 g^2 |\Psi|^2 \Psi^2 H_0^2 \Psi^* - 4 g^2 H_0(|\Psi|^2 \Psi^2 H_0 \Psi^*) \\
& + 2 g^2 |\Psi|^2 H_0^2(|\Psi|^2 \Psi) - g^2 \Psi^2 H_0^2(|\Psi|^2 \Psi^*) + 3 g^2 H_0(|\Psi|^4 H_0 \Psi) \\
& - 2 g^2 |\Psi|^2 H_0 (\Psi^2 H_0 \Psi^*) + 4 g^2 |\Psi|^2 H_0(|\Psi|^2 H_0 \Psi) + 4 g^2 |\Psi|^2 \Psi^* (H_0 \Psi)^2 \\
& + 3 g^2 |\Psi|^4 H_0^2 \Psi + 4 g^2 \Psi^3 (H_0 \Psi^*)^2 - 22 g^2 |\Psi|^2 \Psi (H_0 \Psi)(H_0 \Psi^*) \\
& + g^2 \Psi^2 H_0 ({\Psi^*}^2 H_0 \Psi) - 2 g^2 \Psi^2 H_0 (|\Psi|^2 H_0 \Psi^*) - 2 g^2 \Psi^2 H_0(|\Psi|^2 H_0 \Psi^*) \\
& + g^3 H_0(|\Psi|^6 \Psi) + 2 g^3 |\Psi|^2 H_0(|\Psi|^4 \Psi) + 3 g^3 |\Psi|^4 H_0(|\Psi|^2 \Psi) \\
& - 11 g^3 |\Psi|^4 \Psi^2 (H_0 \Psi^*) + 6 g^3 |\Psi|^2 \Psi^2 H_0(|\Psi|^2 \Psi^*) - g^3 \Psi^2 H_0(|\Psi|^4 \Psi^*) \\
& + 4 g^3 |\Psi|^6 H_0 \Psi + g^4 |\Psi|^8 \Psi.
\end{array}
\label{appeq4}
\end{equation}
\end{widetext}
To compare these with (\ref{appeq1}), we construct the following non-trivial products of $\hat{V}_{nl}$ and $H_0$, using the composition rule (\ref{neweq14}):
\begin{equation}
\hat{V}_{nl} H_0 \Psi = 2 g |\Psi|^2 H_0 \Psi - g \Psi^2 H_0 \Psi^*,
\label{appeq5}
\end{equation}
\begin{widetext}
\begin{equation}
\hat{V}_{nl}^2 H_0 \Psi = 3 g^2 |\Psi|^4 H_0 \Psi - 2 g^2 |\Psi|^2 \Psi^2 H_0 \Psi^*,
\label{appeq6}
\end{equation}
\begin{equation}
\hat{V}_{nl}^3 H_0 \Psi = 4 g^3 |\Psi|^6 H_0 \Psi - 3 g^3 |\Psi|^4 \Psi^2 (H_0 \Psi^*)
\label{appeq7}
\end{equation}
\begin{equation}
\hat{V}_{nl} H_0^2 \Psi = g \Psi^2 H_0^2 \Psi^* - 4 g \Psi (H_0 \Psi) (H_0 \Psi^*) + 2 g \Psi^* (H_0 \Psi)^2 + 2 g |\Psi|^2 H_0^2 \Psi,
\label{appeq8}
\end{equation}
\begin{equation}
\hat{V}_{nl} H_0 \hat{V}_{nl} \Psi = g^2 \Psi^2 H_0 (|\Psi|^2 \Psi^*) - 2 g^2 |\Psi|^2 \Psi^2 (H_0 \Psi^*) + 2 g^2 |\Psi|^2 H_0 (|\Psi|^2 \Psi),
\label{appeq9}
\end{equation}
\begin{equation}
\begin{array}{ll}
\hat{V}_{nl} H_0^3 \Psi = & - g \Psi^2 H_0^3 \Psi^* + 6 g \Psi (H_0 \Psi) (H_0^2 \Psi^*) - 6 g \Psi (H_0 \Psi^*)(H_0^2 \Psi) \\
& + 2 g |\Psi|^2 H_0^3 \Psi - 6 g (H_0 \Psi)^2 (H_0 \Psi^*) + 6 g \Psi^* (H_0 \Psi) (H_0^2 \Psi)
\end{array}
\label{appeq10}
\end{equation}
\begin{equation}
\begin{array}{ll}
\hat{V}_{nl} H_0^2 \hat{V}_{nl} \Psi = & - g^2 \Psi^2 H_0^2(|\Psi|^2 \Psi^*) + 4 g^2 \Psi (H_0 \Psi) H_0(|\Psi|^2 \Psi^*) \\
& + 2 g^2 |\Psi|^2 \Psi^2 H_0^2 \Psi^* - 4 g^2 \Psi (H_0 \Psi^*) H_0 (|\Psi|^2 \Psi) \\
& + 2 g^2 |\Psi|^2 H_0^2 (|\Psi|^2 \Psi) -2 g^2 |\Psi|^2 \Psi^* (H_0 \Psi)^2 \\
& -4 g^2 |\Psi|^2 \Psi (H_0 \Psi)(H_0 \Psi^*) + 4 g^2 \Psi^* (H_0 \Psi) H_0 (|\Psi|^2 \Psi) 
\end{array}
\label{appeq11}
\end{equation}
\begin{equation}
\begin{array}{ll}
\hat{V}_{nl} H_0 \hat{V}_{nl} H_0 \Psi = & -2 g^2 \Psi^2 H_0 (|\Psi|^2 H_0 \Psi^*) + g^2 \Psi^2 H_0({\Psi^*}^2 H_0 \Psi) \\
& + 2 g^2 |\Psi|^2 \Psi^2 H_0^2 \Psi^*  + 2 g^2 \Psi (H_0 \Psi) H_0(|\Psi|^2 \Psi^*) \\
&- 2 g^2 \Psi (H_0 \Psi^*)H_0(|\Psi|^2 \Psi) + 2 g^2 \Psi^3 (H_0 \Psi^*)^2 \\
& + 4 g^2 |\Psi|^2 H_0(|\Psi|^2 H_0 \Psi) - 2 g^2 |\Psi|^2 H_0 (\Psi^2 H_0 \Psi^*) \\
& - 6 g^2 |\Psi|^2 \Psi (H_0 \Psi) (H_0 \Psi^*) + 2 g^2 \Psi^* (H_0 \Psi) H_0 (|\Psi|^2 \Psi) \\
\end{array}
\label{appeq12}
\end{equation}
\begin{equation}
\begin{array}{ll}
\hat{V}_{nl}^2 H_0^2 \Psi = & 2 g^2 \Psi^3 (H_0 \Psi^*)^2 + 2 g^2 |\Psi|^2 \Psi^2 H_0^2 \Psi^* - 12 g^2 |\Psi|^2 \Psi (H_0 \Psi) (H_0 \Psi^*) \\
& + 6 g^2 |\Psi|^2 \Psi^* (H_0\Psi)^2 + 3 g^2 |\Psi|^4 H_0^2 \Psi
\end{array}
\label{appeq13}
\end{equation}
\begin{equation}
\begin{array}{ll}
\hat{V}_{nl} H_0 \hat{V}_{nl}^2 \Psi = & - g^3 \Psi^2 H_0 (|\Psi|^4 \Psi^*) + 4 g^3 |\Psi|^2 \Psi^2 H_0 (|\Psi|^2 \Psi^*) - 4 g^3 |\Psi|^4 \Psi^2 (H_0 \Psi^*) \\
& + 2 g^3 |\Psi|^2 H_0 (|\Psi|^4 \Psi)
\end{array}
\label{appeq14}
\end{equation}
\begin{equation}
\hat{V}_{nl}^2 H_0 \hat{V}_{nl} \Psi = 2 g^3 |\Psi|^2 \Psi^2 H_0 (|\Psi|^2 \Psi^*) - 4 g^3 |\Psi|^4 \Psi^2 H_0 \Psi^* + 3 g^3 |\Psi|^4 H_0(|\Psi|^2 \Psi)
\label{appeq15}
\end{equation}
\end{widetext}
Using (\ref{appeq5})-(\ref{appeq15}) in (\ref{appeq1}), we do indeed reproduce the exact expressions for $\partial_t^n \Psi$ found in (\ref{appeq2})-(\ref{appeq4}).  Thus, we have explicitly shown that, for the Gross-Pitaevskii equation, the definition for $\hat{V}_{nl}$ (\ref{neweq14}) and Taylor expansion of the exponential yields
\begin{equation}
\Psi(t) = \exp\left(-i t (H_0 + \hat{V}_{nl})\right) \Psi(0) + O(t^5).
\end{equation}

\end{document}